\documentclass[usenatbib]{mn2e}

\newcommand{\rxj}{RX~J0806.3+1527}

\usepackage{graphicx}
\usepackage{times}

\title[Constraints upon the unipolar model of  V407~Vul and 
RXJ0806.3+1527]{Geometrical constraints upon the unipolar model of  V407~Vul and 
RXJ0806.3+1527}
\author[S.C.C. Barros et al. ]{S.C.C. Barros$^{1}$\thanks{E-mail:s.c.barros@warwick.ac.uk} 
T.R. Marsh$^{1}$,  P. Groot$^{2}$, G. Nelemans$^{3,2}$, G. Ramsay$^4$,\and
 G. Roelofs$^2$ D. Steeghs$^{5}$ and J. Wilms$^{1}$\\ 
$^{1}$Department of Physics, University of Warwick, Coventry, CV4 7AL, UK\\
$^{2}$University of Nijmegen, P.O. Box 9010, 6500 GL Nijmegen, The Netherlands\\
$^{3}$Institute of Astronomy, University of Cambridge, Madingley Road, Cambridge CB3 OHA, UK\\
$^{4}$Mullard Space Science Laboratory, University College London, Holmbury St.\ Mary, Dorking, UK\\
$^{5}$Harvard-Smithsonian Center for Astrophysics, Cambridge, USA}

\begin{document}
\date{Accepted 1988 December 15. Received 1988 December 14; in original form 
1988 October 11}


\pagerange{\pageref{firstpage}--\pageref{lastpage}} \pubyear{2002}

\maketitle

\label{firstpage}

\begin{abstract}
  V407~Vul and \rxj\ are X-ray emitting stars with X-ray light curves that are
  100\% modulated on periods of 569 and 321 seconds respectively. These periods
  and no others are also seen at optical and infra-red wavelengths. These
  properties have led to the suggestion that the periods are the orbital periods
  of ultra-compact pairs of white dwarfs. There are two main double white dwarf
  models: the unipolar inductor model analogous to the Jupiter--Io system and the
  direct impact model analogous to Algol.  In this paper we consider geometrical
  constraints on the unipolar inductor model, in particular what parameter
  values (component masses, orbital inclination and magnetic co-latitude) can describe the X-ray and
  optical light curves. We find that for a dipole field on the primary star, the
  unipolar inductor model fails to match the data on V407~Vul for any
  combination of parameters, and can only match RX~J0806.3+1527 if the
  sparser set of observations of this star have been unluckily timed.
\end{abstract}


\section{Introduction}
In recent years much attention has been paid to what are possibly the
shortest period binary stars known, V407~Vul ($P = 570$
sec, \citealt{cropper1998a,motch1996a}) and \rxj\  ($P =321$ sec,
\citealt{israel1999a}).
 Both of these stars show highly modulated X-ray light curves
which are on for about 60\% of the time and off for the remaining 40\%
 \citep{cropper1998a, israel1999a}.
Both stars also show only one period (and its harmonics) at all wavelengths observed
\citep{ramsay2000a,ramsay2002b,israel02a}. These and other properties have lead to the suggestion that the periods
may be orbital, only possible for a pair of compact objects, most likely white
dwarfs. This would make these systems strong emitters of gravitational waves and
possible progenitors of the semi-detached AM~CVn stars.

There are several rival models for these systems. V407~Vul was first suggested
to be an intermediate polar (IP), in which case its $9.5$ minute period would be
the spin period of an accreting magnetic white dwarf \citep{motch1996a}.
However, the lack of any other period representative of a longer period binary
orbit lead \citet{cropper1998a} to suggest instead that V407~Vul was a polar
containing a white dwarf with a strong enough magnetic field to lock its spin to
the orbit, making it the first known ``double degenerate'' polar. This model
received a blow when no polarisation was detected \citep{ramsay2000a}, which lead
\citet{Wu02a} to develop a unipolar inductor model. In the unipolar
inductor model a slight
asynchronism between the spin period of a magnetic white dwarf and the orbital
period within a detached double white dwarf binary creates an electric current
between the two components of the binary. The dissipation of this current powers the
observed X-ray flux.  Following \citet{Wu02a}, an alternative double white dwarf
model was developed by \citet{marsh2002a}. They proposed that accretion
could be taking place without a disc \citep{nelemans2001b,ramsay2002a}, since for
periods below $\sim 10$ min and for plausible system parameters it becomes possible for the mass transfer stream to
crash directly onto the accreting white dwarf.
\rxj, although of a shorter period, is so similar to V407~Vul that all the above
models apply equally well to it as they do to V407~Vul.  Finally, the IP model
was resuscitated by \citet{norton04a} with a model in which we see the systems
almost face-on with the X-ray variations caused by the accretion stream flipping
completely from one magnetic pole to the other each cycle.

There is as yet no clear winner out of the various models, all of which face
difficulties. In the direct impact model, we expect the (orbital) period to be
increasing, whereas measurements show it to be decreasing in both
V407~Vul \citep{strohmayer2002a,strohmayer04a} and
\rxj\ \citep{hakala2003a, strohmayer03a, hakala2004a}. The weakness of optical emission lines in
\rxj\ and V407~Vul also
seems hard to reconcile with an accreting binary, even though direct impact may
produce weaker line emission than disc accretion \citep{marsh2002a}.
The weakness of the optical emission lines is probably also the most difficult fact to
accommodate within polar or IP models since all such systems discovered to date
exhibit strong lines in their spectra. The IP model \citep{norton04a} is also
unattractive for the fine-tuning it requires for two independent systems. The
unipolar inductor model faces only a theoretical problem: it is short lived
($\sim 1000$ years) because it lives off the spin energy of one of its two white
dwarfs. As yet, no way of creating/maintaining asynchronism in the face of strong
dissipation has been proposed. It does however nicely match the decreasing
period of both systems, and so can justly be claimed to be the leading model 
at present \citep{hakala2004a}.

Until a better alternative is found, we are faced with trying to select the
``least worst'' amongst the models, and every model must be tested as far as
possible against the available observations. In this paper we examine whether
the light curves are compatible with the unipolar inductor model, an area not
considered in detail by \citet{Wu02a}.  In section~3 we define the
geometry that we use to describe the \citet{Wu02a} model then we present our results
in section~4.  We begin however by reviewing the observational constraints
deduced from the X-ray light curves.

\section{Observational  Constraints}
\label{sec:constraints}
Our aim is to see if the observed light curves have a natural explanation under
the electric star model. In doing so, it is important to avoid fine detail as
probably all models fail at this level. For instance, in its simplest form of
a small spot at the equator, the direct impact model gives 50:50 on/off, not
the observed 60:40 ratio. However, it is not impossible to imagine that
vertical structure near the impact site along with heating spreading downstream
from it could lead to an increase of the ``on'' phase. Thus we try to distill
key features which should be explained on any model, without being over specific
in our selection. We identify the following key constraints:
\begin{enumerate}
  
\item[I.] The bright phase during which X-rays are seen from the heated
  spots has to \textit{last for less than 0.6 of the cycle}. While, as
  described above, it is possible to extend the visibility period of a given
  simple model, it is much harder to decrease it, so we believe that this is a
  fundamental restriction upon any model.
  
\item[II.] The bright phase has to be \textit{more than 0.4 of the cycle} because the
  maximum size of a spot (see later) extends the visibility period by 0.2 of a
  cycle at most. 

\item[III.] The phase of the X-ray maxima must deviate from a smooth long-term trend
  by at most $0.2$ of a cycle, in order to match the observations
  \citep{strohmayer04a,strohmayer03a}.
  
\item[IV.] The light curves show no hint of eclipses and so we take the orbital
  inclination angle $i$ to satisfy:
\begin{equation}
\cos i  \ge \frac{r_1+r_2}{a}
\end{equation}
where $r_1$ and $r_2$ are the radii of the two stars and $a$ is the binary separation.
\end{enumerate}

\section{The Geometric Model}

\begin{figure}\centering
\includegraphics[width=\columnwidth]{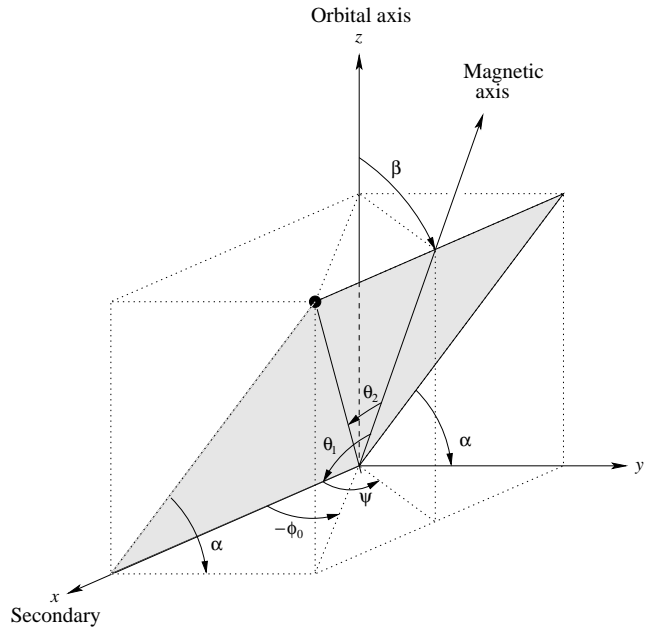}
\caption{Orientation of the primary
  relative to the orbital axis and  direction of the secondary centre. We
  also define the orientation of the magnetic field of the primary, tilted by
  $\beta$ from the $z$ axis and whose projection in $xy$ plane makes an angle
  $\psi$ with the $x$-axis. The dot represents where the field line that goes
  through the secondary crosses the surface of the primary, the position of the
  spot makes an angle $\theta_1-\theta_2$ with the $x$-axis.}
\label{fig:mangle}
\end{figure}

Following the model presented by \citet{Wu02a} we consider a system in which one
star (which we call the secondary star) orbits within the magnetic field of its
companion, the primary star. An asynchronism between the spin of the primary
star and the orbit induces an electric current to flow along the field lines
connecting the two stars, with energy dissipated mainly where these field lines
enter the surface of the primary star. In the Jupiter--Io system the observed
emission lies within $6^\circ$ of the theoretically predicted foot point
\citep{clarke1996a}. In our case we expect the emission to be even
nearer to the foot point
because of the smaller degree of asynchronism.

Our task is to identify the positions of these foot points,
and then work out for how long they are visible for a given orbital inclination.
We treat them as points in order to allow the model to match the 40\% dark phase
during which X-rays disappear (constraint I, section~\ref{sec:constraints}) as
easily as possible.  To allow the model flexibility in matching the bright
phase, we allow for the maximum size of the spots, as described at the end of
this section.

We will now describe the geometry of the model with reference to
Fig.~\ref{fig:mangle}.  We work with a right-handed Cartesian coordinate system
$(x$,$y$,$z)$ centred upon the primary star, with the $x$-axis pointing towards
the secondary star, the $y$-axis parallel to the direction of motion of the
secondary star, and the $z$-axis along the orbital axis (Fig.~\ref{fig:mangle}). The
axis of the magnetic dipole is tilted by an angle $\beta$ (often called the
magnetic colatitude) with respect to the spin and orbital axes, which
we take to be aligned. The azimuthal angle or magnetic
longitude of the dipole $\psi$ is defined by the angle between its projection
onto the orbital plane (the $x$-$y$-plane) and the $x$-axis. In
\citet{Wu02a}'s model there is an asynchronism between the period of the
binary and the spin period of the magnetic white dwarf. Therefore $\psi$ is
the angle that completes a cycle every few days owing to the asynchronism
between the orbital frequency $\omega_o$ and the spin frequency of the primary
star $\omega_s$:
\begin{equation}
\psi=(\omega_s-\omega_o)t+\psi_0.
\end{equation}

 \citet{Wu02a} estimate the asynchronism $(\omega_s-\omega_o)/\omega_o$ to be  1 part
in 1000. This means that all possible azimuths of the magnetic axis compared to
the line of centres of the two stars are explored within $\sim 1000$ orbits,
which is only a few days. If the magnetic axis is tilted with respect to the
orbital axis, such azimuth variation causes the phase of the light curve maxima
to vary back and forth relative to any long term trend.  By applying
the III condition upon the phase, we effectively assume that the X-ray observations have
sampled all possible relative orientations of the magnetic axis (all values of $\psi$ have
been explored) and therefore we insist that the constraints of the previous section are satisfied
for all $\psi$. Ideally this
would be the case, and indeed with sufficient observations one can get very close
to it. We will consider how close we are to this in reality in
section~\ref{sec:phases}.

We define the positions of the foot points by where the magnetic field line
which passes through the centre of the secondary star crosses the surface of the
primary star. This field line lies in a plane defined by the $x$-axis and the
magnetic dipole vector which makes an angle $\alpha$ with the orbital plane Fig.~\ref{fig:mangle}.
We define the orientation of the foot points in this plane with two angles 
$\theta_1$ and $\theta_2$ which measure the angle between the dipole axis and $x$-axis and
the angle between the dipole axis and the foot point respectively. 

Using the equation for dipole field lines $r=\mathcal{C}\sin^2\theta$, we can deduce the following relation between $\theta_1$ and $\theta_2$
\begin{equation}
\sin\theta_2 = \sqrt\frac{r_1}{a}\sin\theta_1 .
\end{equation}
while from simple geometry one can show that
\begin{equation}
\cos \theta_1 = \cos \psi \sin \beta,
\end{equation}
and that
\begin{equation}
\cos \alpha = \frac{\sin \psi \sin \beta}{\sqrt{(\sin \psi \sin\beta)^2 + \cos^2\beta}}.
\end{equation}

To be able to calculate the visibility of the spot from the Earth, we need the orbital
phase $\phi$ and the orbital inclination angle $i$, which we define in
the standard manner, i.e. $i = 90^\circ$ for an edge-on eclipsing
system,  $\phi = 0^\circ$ when  the magnetic primary star is at its furthest from Earth.
With all angles defined, a given spot is visible while
\begin{equation}
\cos(\phi-\phi_0) > -\frac{h}{A}
\end{equation}
where
\begin{equation}
A =\sin i\sqrt{\cos^2(\theta_1-\theta_2) + \cos^2\alpha\,\sin^2(\theta_1-\theta_2)},
\end{equation}
\begin{equation}
h = \cos i \sin \alpha \sin(\theta_1-\theta_2),
\end{equation}
and
\begin{equation}
\tan \phi_0 = \frac{-\cos \alpha \, \sin(\theta_1-\theta_2)}{\cos(\theta_1-\theta_2)}.
\end{equation}
For given values of $\beta$, $i$, $\psi$ and $r_1/a$, it is a straightforward matter
to calculate the remaining variables, and in particular to calculate the fraction of the
cycle over which the spot can be seen. Before describing these calculations, we pause
to consider the issue of the size of the foot points, which is important
for the second of our constraints which defines the minimum period over which the spots are
visible.

The sizes of the foot points are related to the size of the secondary star. If
the latter is tiny, then so too are the foot points. \citet{Wu02a} give an
equation for the size of the foot points (A1 of their appendix A) from which we
deduce that the maximum extent in longitude of the foot points, is of order $2
r_2/a$ radians.  This will lengthen the bright phase by a corresponding amount.
The maximum extent is thus set by the maximum relative size of the secondary,
which is itself set by Roche geometry. For equal mass ratios, we deduce a
maximum longitude extent of about $40^\circ$, which could extend the bright
phase by $\sim 10$\% of the cycle. To be conservative, we assume a maximum
lengthening of $0.2$ for all models, as outlined in constraint II of section~\ref{sec:constraints}.

In our model we assume that the X-rays are only emitted at one of the two foot
points. Allowing both spots to contribute makes matters worse. Although the 
restriction that the bright phase last for more than 40\% of the cycle becomes 
easier to satisfy, it is more difficult to fit the 40\% dark interval. The phase restriction is also affected
negatively: for most configurations only one spot would be visible at a given
time, but when the two spots are visible there would be a change in $\psi$
leading to a phase shift.

\subsection{Computations}
We computed the fraction of the cycle that one of the foot points
is visible as a function of magnetic colatitude $\beta$ (from 0 to $180^\circ$)
and orbital inclination (0 to $90^\circ$). In order to implement our
assumption that we have observed all possible values of $\psi$, we 
search for the maximum and minimum values of the spot visibility over
a finely-spaced array of $\psi$ values from 0 to $180^\circ$ (symmetry allows us to 
avoid searching over $360^\circ$). In order to
obey the constraints of section~\ref{sec:constraints}, we must have that
\begin{enumerate}
\item the \emph{maximum} time that the spot is visible must be \emph{less} than 0.6 of a cycle,
\item the \emph{minimum} time that the spot is visible must be \emph{more} than 0.4 of a cycle.
\end{enumerate}
The masses of the white dwarfs have to be less than $1.4 \,\mathrm{M}_\odot$ 
(Chandrasekhar's limit) and larger than the Roche lobe filling mass: for  
RXJ0806.3+1527  $M_{1},M_{2} > 0.12\,\mathrm{M}_\odot$ and for V407 Vul $M_{1},M_{2} > 
0.065\,\mathrm{M}_\odot$. We calculated the binary separation using Kepler's third law and the radius 
of the white dwarf using Eggleton's mass-radius relation quoted by \citet{verbunt1988a}.

\section{Results}

\begin{figure*}
\hspace*{\fill}
\includegraphics[width=0.9\columnwidth]{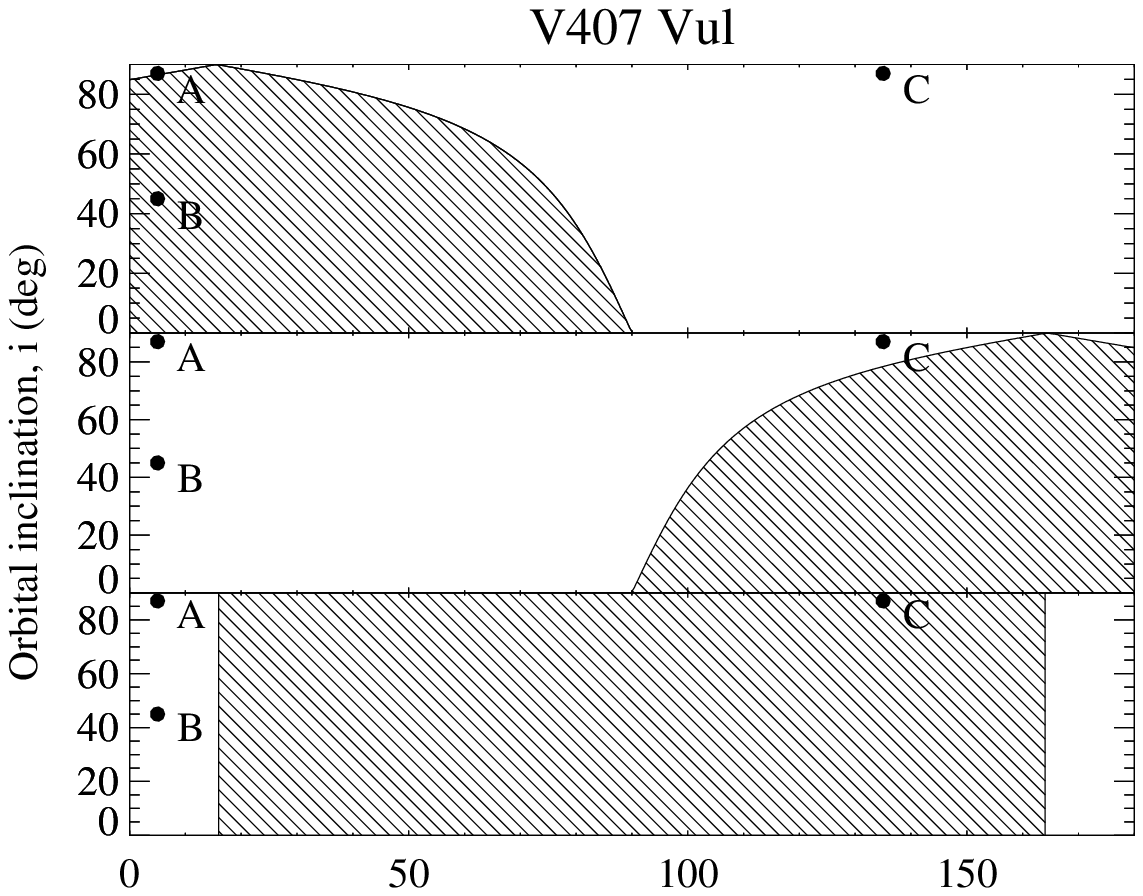}
\hspace*{\fill}
\includegraphics[width=0.9\columnwidth]{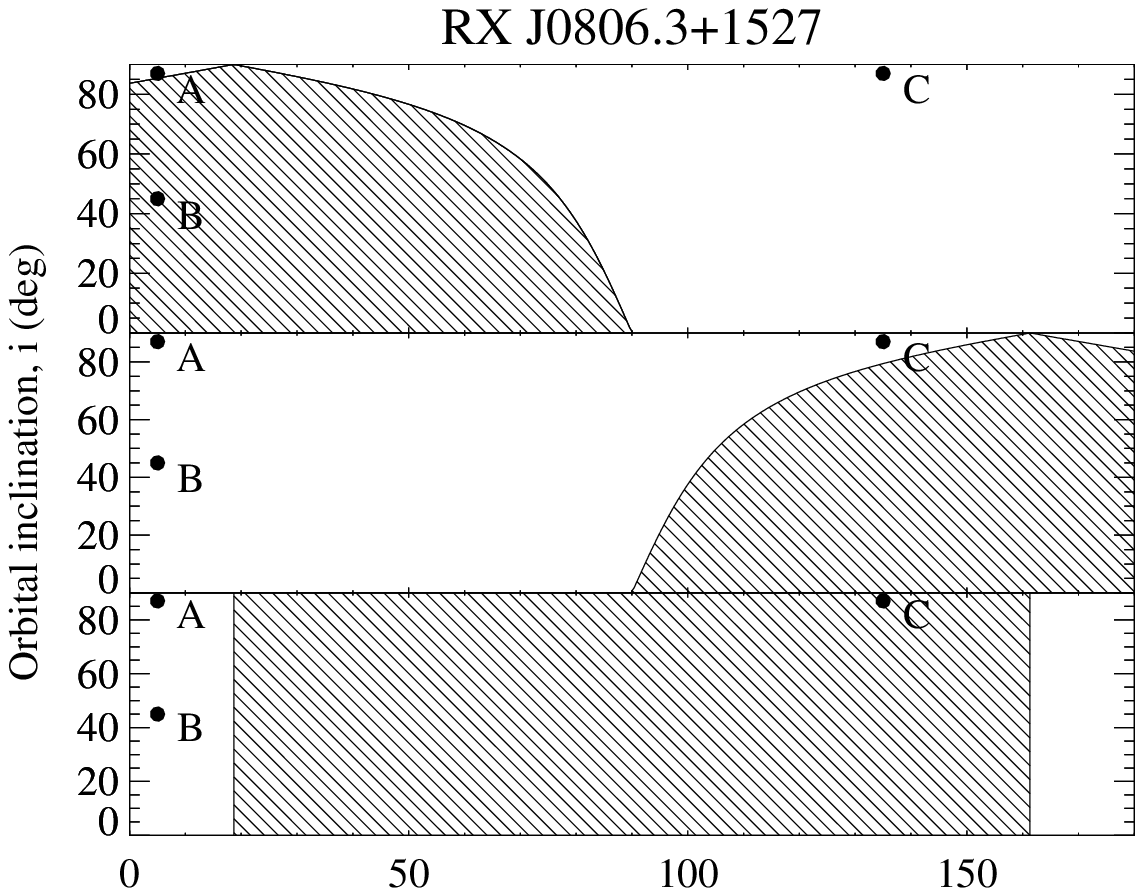}
\hspace*{\fill}
\caption{ Geometrical constraints for the unipolar modal, for V407~Vul
  in the left hand panel and for RXJ0806.3+1527 in the right hand
  panel. The shaded areas represent the parameter
  space ruled out by the restrictions I, II, III respectively in the top,
  middle and bottom panels.}
\label{fig:cons6}
\end{figure*}

\begin{figure}\centering
\includegraphics[width=\columnwidth]{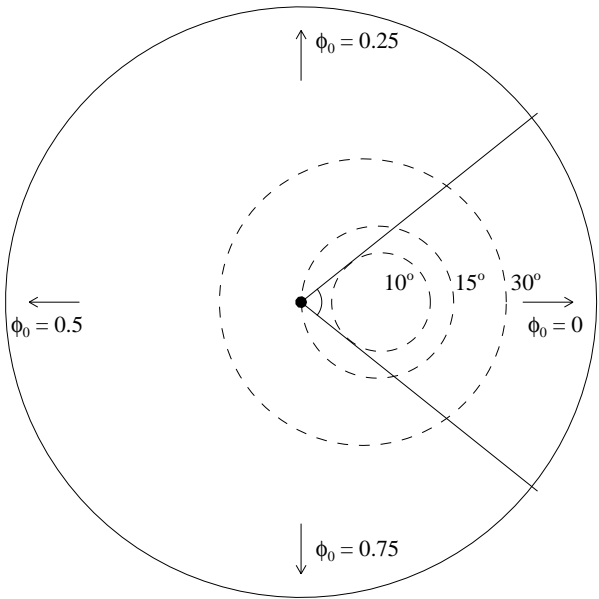}
\caption{Track of the heated spot (dashed lines) viewed from
above the primary star (outermost circle) for different choices of inclination of the magnetic
field $\beta$. For $\beta$ smaller than the critical
angle $\beta_c$ the track is always on the side facing the secondary (towards
the right). In this case the X-ray pulse phase varies over a
restricted range indicated here by the two tangential lines for
$\beta=10^\circ$. For $\beta$ greater than $\beta_c$ the heated spot makes a complete circle
around the polar axis of the primary star causing a complete phase shift of the X-ray pulses over one beat
period.}  
\label{fig:path}
\end{figure}

\begin{figure}\centering
\includegraphics[width=\columnwidth]{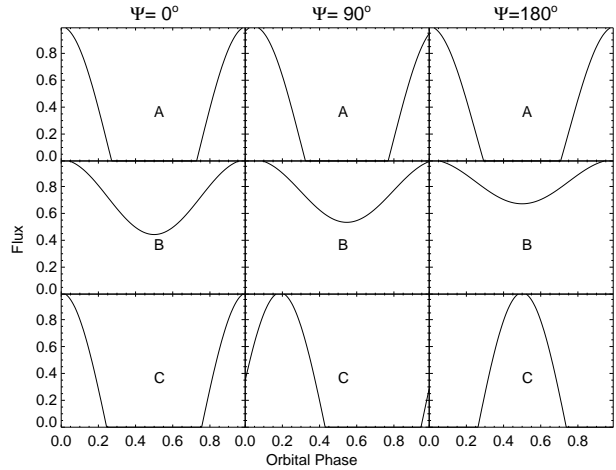}
\caption{The figure shows examples of X-ray phase light curves
for  V407~Vul.  A, B and C represent different choice of parameters
shown in Fig.~\ref{fig:cons6} with three possible $\psi=$\  $0^\circ$, $90^\circ$ and
$180^\circ$ to illustrate the phase shift. For A,  $\beta=5^\circ$ and
$i=87^\circ$, for B, $\beta=5^\circ$ and $i=45^\circ$ and for C,
$\beta= 135^\circ$ and $i=87^\circ$. In all the figures the flux is normalised to unity.}
 \label{fig:vlight}
\end{figure}

We start by discussing how the constraints of section~\ref{sec:constraints} restrict the
possible values of the magnetic tilt $\beta$ and the inclination angle
$i$  for typical white dwarf masses of  $0.6\,\mathrm{M}_\odot$
(Fig.~\ref{fig:cons6}). The first constraint (top panels) rules
out combinations of low inclination and small angle between the magnetic field
and the spin axis.  This is because we would be looking at the north pole and
the spot would also be in the northern hemisphere, i.e. it would be visible for
most of the time which would violate the maximum 60\% bright phase.
Note that there is a critical magnetic colatitude when $\beta_c = \theta_2$
given by
\begin{equation}
\tan \beta_c = \sqrt{r_1/a}
\end{equation}
such that when the magnetic axis is tilted away from the secondary star ($\psi = 180^\circ$),
the foot point lands upon the spin pole of the primary star and will therefore always be visible, 
no matter the orbital inclination. For $M_1 = M_2 = 0.6\,\mathrm{M}_\odot$, $\beta_c = 15^\circ$ for 
V407~Vul and $19^\circ$ for \rxj. This critical value is clear in
Fig.~\ref{fig:cons6}.

The second constraint is symmetric to the first, so that the foot point cannot
lie too close to the south pole without leading to too short a visibility
fraction (central panels of Fig.~\ref{fig:cons6}).

The third constraint, that of phase, is highly restrictive as it removes all
colatitude values in the range $\beta_c < \beta < 180 - \beta_c$. In other
words, the magnetic and orbital axes must be nearly aligned in order for the
phases not to wander. The reason for this is that for $\beta$ in the excluded
range, the phase switches by 180 degrees depending upon whether the magnetic
dipole is oriented towards or away from the secondary star.  This would show
up as a large deviation in the X-ray pulses from a smooth trend.
Note that our exact phase constraint ($< 0.2$ of a cycle variation) is actually
a little more restrictive still since significant phase wander occurs as 
$\beta$ nears $\beta_c$.

To better illustrate this, Fig.~\ref{fig:path} shows the track of
the heated spot on the primary star during one beat period as seen in the
rotating frame of the binary. We can see that,
for  $ \beta < \beta_c$ (or the symmetric case $\beta > 180 - \beta_c$), the phase offset is restricted
to lie within the region defined by the two tangential lines to the small dashed circle
but for $\beta >  \beta_c $ the phase offset goes through a complete
cycle. For the dipole field we assume, one can show that the tracks are perfect circles.

To visualise the three constraints Fig.~\ref{fig:vlight} shows examples of
the light curves using some combinations of parameters that
illustrate the problems of this geometry.
When $\beta=5^\circ$ and $i=87^\circ$ (marked `A' on Fig.~\ref{fig:cons6}) the light
curve obeys the first three constraints and it is similar to the
observed data. In the second case considered $\beta=5^\circ$ and $i=45^\circ$
(`B' in Fig.~\ref{fig:cons6}) the spot is always visible and therefore fails to obey constraint
I. In the third example $\beta= 135^\circ$ and $i=87^\circ$ (`C')
the phase constraint is not complied with and there is a large phase shift
between $\psi=0^\circ$, and
$\psi=180^\circ$.

The fourth and last constraint is the restriction that the system
does not eclipse. This is independent of magnetic tilt and depends
only upon the masses of the two stars for a given orbital period.

\begin{figure*}
\hspace*{\fill}
\includegraphics[width=0.9\columnwidth]{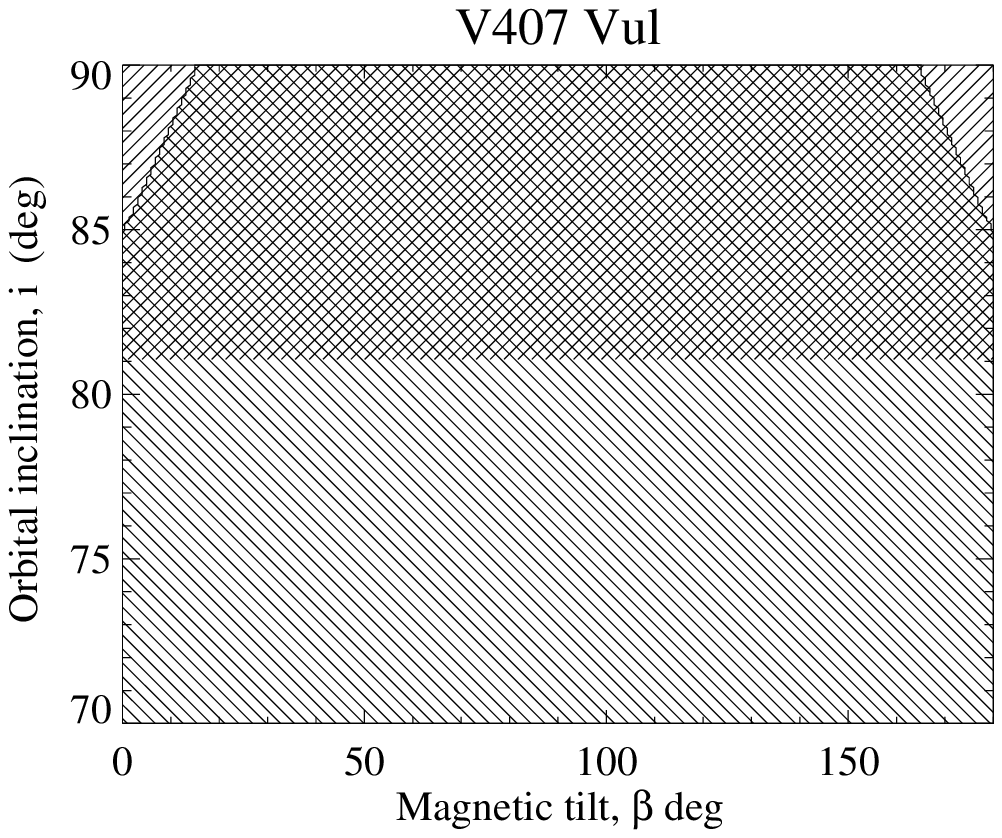}
\hspace*{\fill}
\includegraphics[width=0.9\columnwidth]{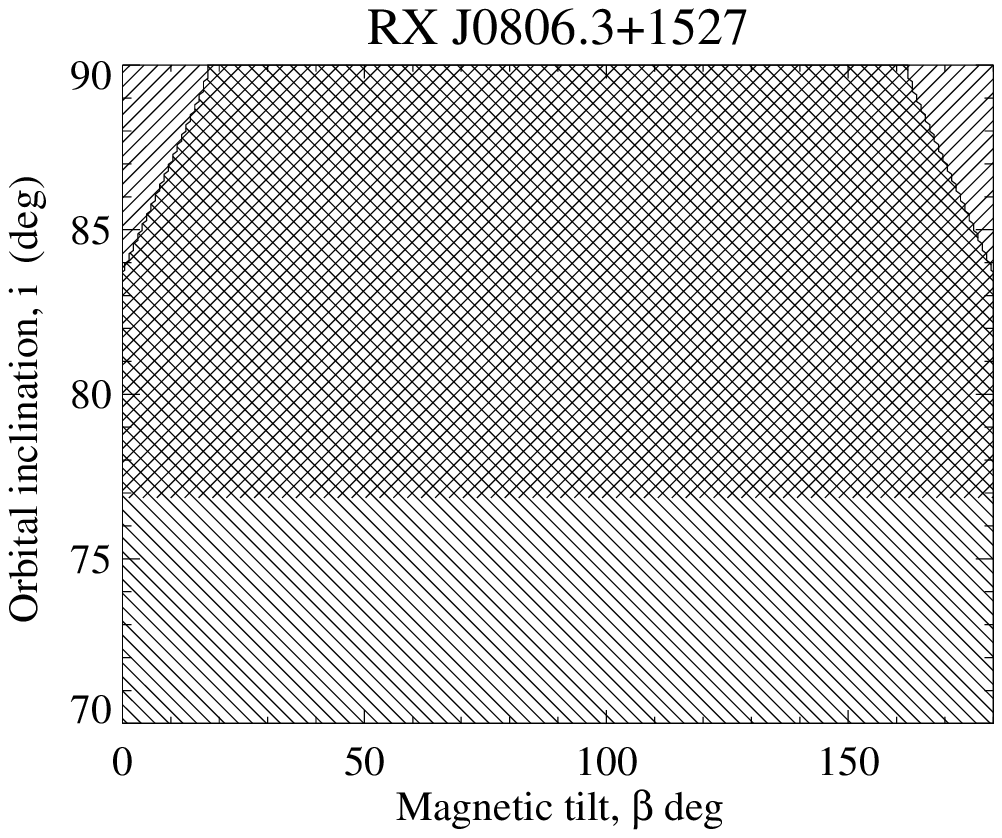}
\hspace*{\fill}
\caption{The lines going from top-left to bottom-right represent the
  parameter space ruled out by the
  combination of the constraints I, II and III. The only regions left out
  are two triangles in the top corners of the parameter space. However these
  are also ruled out when we  include  the absence of eclipses which
  rules out the regions shown
  by the oppositely sloped lines. Therefore we are 
  left with no possible solution in the case of  $M_1=0.6M_\odot$ and
  $M_2 = 0.6M_\odot$. Note that the vertical scale here has been adjusted to
  show only the top parts of Fig.~2}
\label{fig:all6}
\end{figure*}

In figure \ref{fig:all6} we show the possible values allowed by the combination
of all the constraints mentioned above, for $M_1=M_2=0.6\,\mathrm{M}_\odot$,
other white dwarf masses will be considered later. The figures also show the
maximum inclination for which the system will not have eclipses (horizontal
lines). Together, the constraints rule out all parameter combinations.

\subsection{Detailed analysis of the phase constraint}
\label{sec:phases}

We assumed for simplicity that we have seen all possible values of the angle $\psi$ which
measures the azimuth of the magnetic dipole within the binary. However, in this
section we investigate further if this is really the case or if the observations
of the two systems carried out so far are not sufficient to eliminate the
possibility that the systems were only observed over a small range of $\psi$,
thereby increasing the available region of parameter space.

We only used the X-ray data because according to the unipolar inductor model,
the optical flux is from the secondary star and is therefore locked to
the orbit.

\citet{Wu02a} deduced that for V407~Vul an asynchronism of 1 part in 1000 would
be enough to explain the luminosity observed.

The degree of asynchronism is a little arbitrary so we explored several
values in the range 1 part in 100000 to 1 part in 700 for V407~Vul. We used the dates and durations of the
X-ray observations of V407~Vul to compute the phase shifts that would have been
seen for a set of degrees of asynchronism in this range.
  The largest asynchronism of 1 part in 700 is set by the length of the longest observation and
is such that this observation covered an entire beat cycle. A degree
of asynchronism larger than
this returns us to our original assumption of complete coverage of
$\psi$. The phase shifts were then fitted with a quadratic ephemeris and the
scatter around the ephemeris was minimised by subtraction or addition of whole
cycles where necessary. This process simulates the treatment of the observed
pulse times by, for example, \citet{strohmayer04a}, and acts to reduce the phase
shifts somewhat. Finally the values were minimised over the (unknown) phase
offset $\psi_0$ by repetition of the computation for 40 values equally spaced
around a cycle and retention of the minimum shift calculated. This is in the
spirit of trying to give the unipolar inductor model the benefit of the doubt
where possible.

We applied the same method to \rxj. In this case
the observations were much shorter and so our range of the degree of
asynchronism is 1 part in 100000 to 1 part in 80 which
corresponds to the period of the longest observation which was made with \textsl{XMM-Newton}.
We reduced archival \textsl{XMM-Newton} data from November 2002 to complement the phase timing
by \citet{strohmayer03a}.  Using the latest published value for the period $P$
and its derivative $\dot{P}$ \citep{hakala2004a} and the phase 0 of the phase
folded X-ray light curve of \citet{israel2003a} $\mathrm{MJD} = 52225.36153 \pm
0.00004$ for comparison, we obtained a phase shift for Nov 2002 close to zero,
so we include the times for these observations in our study. However, given
present uncertainties in both $P$ and $\dot{P}$, we cannot rule out a larger
phase shift between any of the observations.

\begin{figure*}
\hspace*{\fill}
\includegraphics[width=0.9\columnwidth]{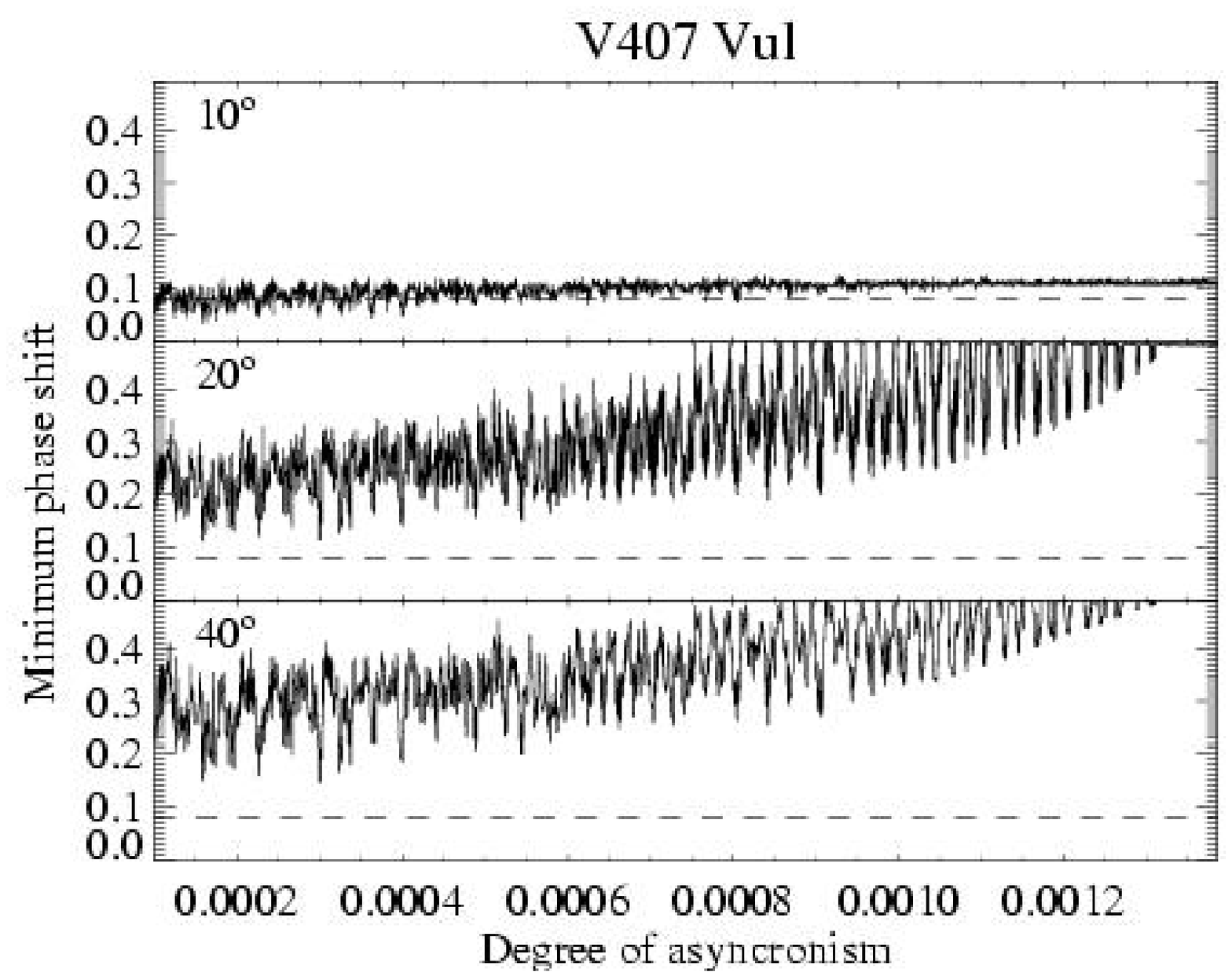}
\hspace*{\fill}
\includegraphics[width=0.9\columnwidth]{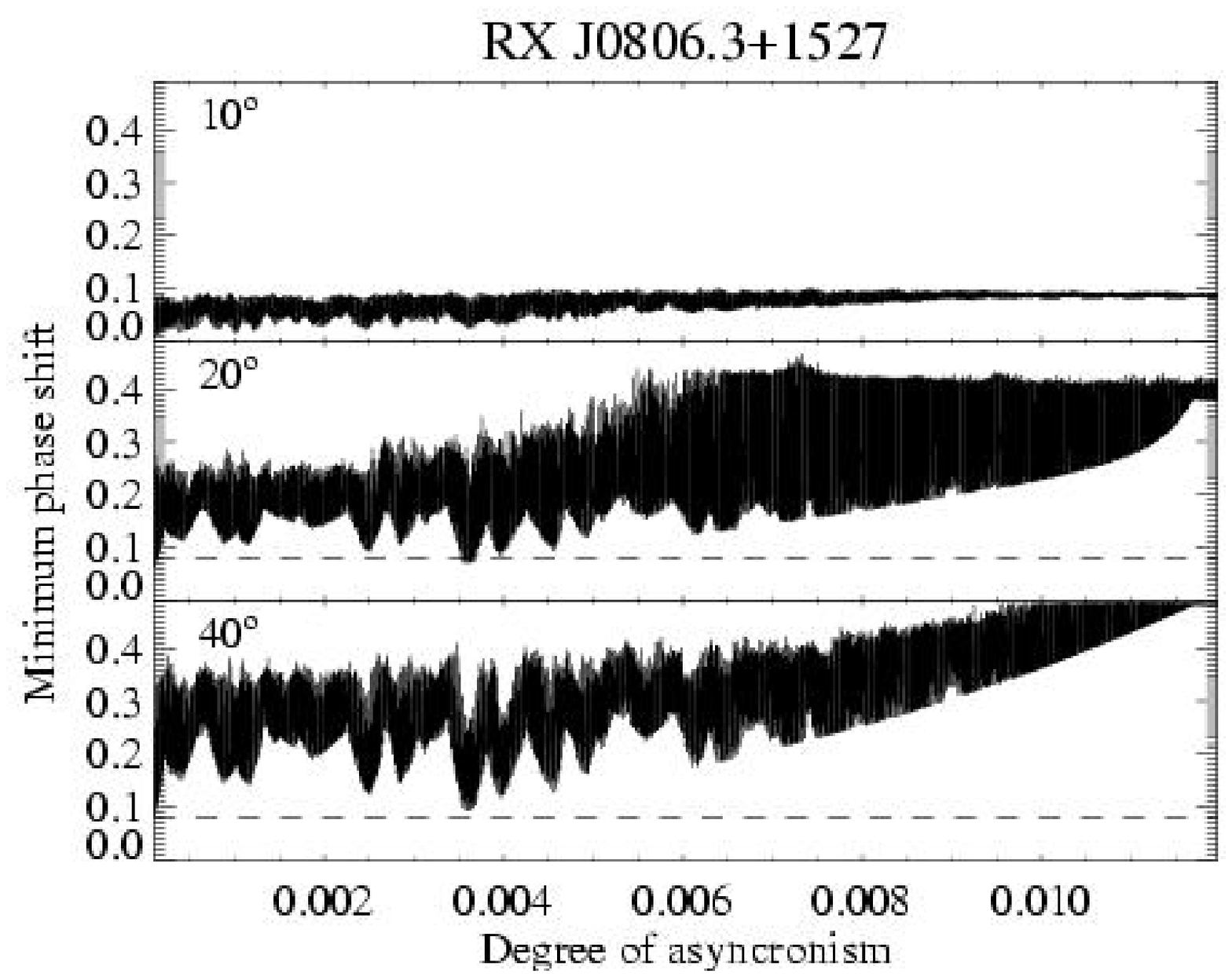}
\hspace*{\fill}

\caption{The minimum phase shift relative to a quadratic ephemeris as
  a function of the degree of spin/orbit asynchronism 
   for $\beta=10^\circ$, $20^\circ$ and $40^\circ$. The dashed line represents the maximum
  phase shift measured in the observations of the two stars $=0.08$.}
 \label{fig:final}
\end{figure*}

Our results are shown in Fig.~\ref{fig:final} in which we plot the phase shift,
minimised over $\psi_0$, as a function of the degree of asynchronism that would have been
seen in the past observations of V407~Vul and \rxj, for different inclinations
of the magnetic field $\beta$ if the \citet{Wu02a} model is correct. (Note that
the phase shift is independent of the orbital inclination $i$.)

As expected from our simple phase constraint, small angles of $\beta$ such that
$\beta < \beta_c$ do have low enough phase shifts to match the observations
which show at most a deviation of $0.08$ cycles for both systems. For V407~Vul,
when $\beta > \beta_c$, even the minimised values that we plot are
consistently larger than is observed, although the phase shifts do not
match the value of $0.5$ that we would predict for perfect observational coverage.
In other words, the observations of V407~Vul are extensive enough that
our conclusion that the region $\beta_c < \beta < 180 - \beta_c$ is ruled out by
the absence of phase shifts remains unchanged.

The much sparser observations of \rxj\ on the other hand, do not allow us to rule
out large $\beta$ angles with confidence as some frequencies
show relatively small phase shifts. Clearly more observations of \rxj\ are
required. Having said this, if the unipolar inductor model were to be let off
the hook by this means, it would be at a significant price because then the
X-ray phases measured so far would not necessarily be a true reflection of the
orbital phase and the present indications of spin-up \cite{hakala2004a}, which
depend in large part upon ROSAT data, could well be spurious.

\section{Discussion}

In this section we will discuss the limitations of our analysis, the
effect of varying the white dwarf masses and the effect of the
spot size upon the allowed parameter space.

\subsection{Effects of different masses}
All of the constraints depend upon the masses of the white dwarfs, so in this
section we explore whether changing the mass can allow some models to work.
Increasing the size of the primary star helps by shifting the foot points toward
its equator. Therefore, we expect an easing of the problem for small primary
star masses, except that this will make the eclipse restriction harder to
satisfy. Increasing the mass of the secondary star on the other hand is entirely
beneficial since it increases the maximum inclination angle for which there is
still no eclipse. 
However, it has to be said that there is some inconsistency
here with respect to the foot point size which will be nowhere near our assumed
maximum. If this could be taken into account reliably, it would count against
large masses for the secondary star, but we do not attempt to allow for this subtlety
here.

For V407~Vul we find that a very small area of the $\beta$--$i$
parameter space is allowed for 
 $M_1 = 1.4 \,\mathrm{M}_\odot$ and $M_2 > 1.3\,\mathrm{M}_\odot$, or
 for $M_1 = M_2 = 1.3\,\mathrm{M}_\odot$. The
parameter space allowed requires the magnetic and spin axes to be nearly aligned
($\beta < 2^\circ$ or $ \beta > 178^\circ $), the system to have a very specific orbital inclination
($88.2 < i < 88.8^\circ$) and masses right at the top of the range for white dwarfs.
Together the combination is very unlikely, and we will see below that it causes
other problems.  The case of \rxj\ is still tougher, and there are only possible
solutions for $M_1=1.4 \, \mathrm{M}_\odot$ and $M_2 =1.4
\,\mathrm{M}_\odot$ which are even more restrictive for 
both $\beta$ and $i$ ($\beta < 1.5^\circ$ or $ \beta > 178.5^\circ $
and $87.8 < i < 88.3^\circ$ )

There are no known white dwarfs with masses as high as required here. Moreover,
even if the masses really were this high, they would be in severe conflict with one
of the main pieces of evidence in favour of the unipolar model, which are the 
measurements of decreasing period, presumed to be caused by gravitational radiation.  We
computed the values of the spin up of the systems for the lowest masses that
have any available parameter space assuming a detached binary driven only by
gravitational radiation loss.  For the case of V407~Vul with $M_1=M_2 =1.3
\,\mathrm{M}_\odot$ the spin up rate would be $\sim 3.6 \times 10^{-16}
\,\mathrm{Hz}\,\mathrm{sec}^{-1}$, some fifty times the measured value reported by
\citet{strohmayer04a} of $7.0 \pm 0.8 \times 10^{-18}
\,\mathrm{Hz}\,\mathrm{sec}^{-1}$, while in the case of \rxj\ for
$M_1=M_2=1.4\,\mathrm{M}_\odot$ the spin up rate would be $\sim 2.3 \times
10^{-15} \,\mathrm{Hz}\,\mathrm{sec}^{-1}$, five times the measured value
reported by \citet{strohmayer03a} and \cite{hakala2004a} of $6.00 \pm 0.1\times 10^{-16}
\,\mathrm{Hz}\,\mathrm{sec}^{-1}$. In consequence even the tiny region
of parameter space that is opened up at high mass fails to match up against what we know of these systems.

\subsection{Spot Size}

We assumed a maximum spot size that would extend the visibility of
a point spot for 0.2 of a cycle which corresponds to the
secondary star filling its Roche lobe , i.e. a very low mass secondary. But
as seen in our analysis the problems facing the unipolar modal are
eased at high masses. As the spot size depends greatly on the secondary mass, the actual
size of the footprints for the masses considered before would be only
0.02 of a cycle for  $M_1 = 0.6 \,\mathrm{M}_\odot$ and 0.003 for
$M_1 = 1.4 \,\mathrm{M}_\odot$ in the case of V407~Vul.
With such a small spot size it would be impossible to accommodate the
model because for example if we force the bright phase to be more than
0.5 of a cycle, we eliminate all the values that were allowed by
constraint I i.e. the first two constraints alone would rule out the model.

\begin{figure}\centering
\includegraphics[width=\columnwidth]{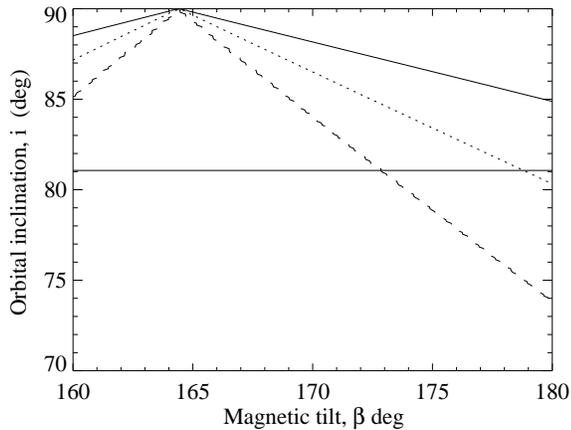}
\caption{The top-left region of the parameter space where we show the
  constraint II for several maximum spot sizes 0.2
  of a cycle (solid line), 0.3 of a cycle (dotted line) 0.6 of a cycle
  (dashed line) for white dwarf masses of  $0.6\,\mathrm{M}_\odot$
  for 407~Vul. The allowed region is always above the line. So for the maximum spot size larger then 0.3 of a cycle there is
 a small allowed region in parameter space (triangle covering  $\beta
 > 173^\circ$ $74^\circ < i < 81^\circ)$.}
\label{fig:size}
\end{figure}

There are some effects that might help make the spot larger, for
example the X-rays could perhaps come from a
vertical extended region (along the field lines), that would allow us to see
them for a little bit longer, although this might be difficult to
reconcile with the thermal nature of the X-ray spectrum. Perhaps the
foot points can leave a trail along the azimuthal direction ($\phi_0$), that would enlarge
the size of the emitting region. The theoretical estimates of the cooling time scale indicate that trailing
is unlikely to be significant \citep{stockman1994a}, but we can in any
case show that increasing the visibility duration has only a limited effect in
helping the unipolar model. In  Fig.~\ref{fig:size} we show the effect
of increasing the spot size upon the available parameter space.
Only for very large spot sizes, covering $> 0.3$ of a cycle in
azimuth, is possible parameter space opened up. (Note that $0.6$ is the
largest possible azimuth extent given the need for a $40\%$ off period.)

In summary, even a fairly radical alteration to allow much larger
spots than are predicted by the unipolar model at best opens up a
small region of parameter space with $\beta > 173^\circ$,  $74^\circ < i
< 81^\circ$. It is clearly unlikely that these applies to two
independent systems.

\section{Conclusion}
Assuming a dipolar field geometry and that the spin and orbital axis
are aligned, we find that the unipolar inductor model presented by \citet{Wu02a} cannot match
the X-ray light-curves of the two systems V407~Vul and \rxj. It fails because
the foot points which produce the X-rays are situated quite close to the magnetic
poles, while at the same time the magnetic axes are forced to lie almost
parallel to the spin axes of the primary stars to avoid excessive phase shifts
in the X-ray pulses. This then means that to obtain the near 50:50 on/off light
curves observed, both systems have to be seen at such high orbital inclination
($> 88^\circ$) that they would eclipse, and yet no eclipses are seen. High
masses for the component stars permit tiny regions of viable orbital
inclination/magnetic inclination parameter space, but lead to very large
gravitational wave losses inconsistent with observed period changes.  
If for some reason the heated spots are much larger than the unipolar
model predicts a small region of parameter space is allowed, but
even this requires fine tuning for two systems.
The only
remaining chinks of light for the unipolar inductor model is that \rxj, owing to
relatively sparse X-ray coverage, could have a relatively highly inclined dipole
without our having spotted the large phase shifts so far, or possibly
the field configurations are very different from a dipole in the two stars. Further observations
are encouraged to close this loophole. We have been careful where possible to
err in favour of the unipolar inductor model; we still have to conclude that it is not viable
in its current form.

\section{Acknowledgements}
SCC Barros is supported by Funda\c{c}\~{a}o para a Ci\^{e}ncia e
Tecnologia e Fundo Social Europeu no \^{a}mbito do III Quadro
Comunit\'{a}rio de Apoio. TRM acknowledges the support of a PPARC
Senior Research Fellowship. GN is supported by NWO-VENI grant 639.041.405. DS acknowledges support of a Smithsonian Astrophysical Observatory Clay Fellowship.

\bibliographystyle{mn2e}
\bibliography{barros}

\label{lastpage}

\end{document}